\documentclass[twocolumn]{emulateapj}

\usepackage{amsmath}
\usepackage{color}

\begin{document}

\title{Electron acceleration at pulsar wind termination shocks}

\author{S. Giacch\`{e} and John G. Kirk}
\affil{Max-Planck-Institut f\"{u}r Kernphysik, Postfach 10 39 80, 69029 Heidelberg, Germany}

\begin{abstract}
\noindent

We study the acceleration of electrons and positrons at an electromagnetically modified, ultra-relativistic shock in the 
context of pulsar wind nebulae (PWNe). 
We simulate the outflow produced by an obliquely rotating pulsar in proximity of its 
termination shock with a two-fluid code which uses a magnetic shear wave to mimic the properties
of the wind. We integrate electron trajectories in the test-particle limit in the resulting background electromagnetic fields
to analyse the injection mechanism. 
We find that the shock-precursor structure energizes and reflects a sizeable fraction of particles, which becomes
available for further acceleration.
We investigate the subsequent first-order Fermi process sustained by small-scale magnetic fluctuations with a Monte Carlo code. 
We find that the acceleration proceeds in two distinct regimes: when the gyro-radius $r_{\textrm{g}}$ exceeds the 
wavelength of the shear $\lambda$, the process is remarkably similar to first-order Fermi acceleration at relativistic,
parallel shocks. This regime corresponds to a low density wind which allows the propagation of superluminal waves.
When $r_{\textrm{g}}<\lambda$, which corresponds to the scenario of driven reconnection, the spectrum is softer.

\end{abstract}


\section{Introduction}

In this paper we address the mechanisms by which relativistic particles are
accelerated in pulsar wind nebulae (PWNe).  The broad-band non-thermal
radiation emitted by these objects is normally modeled as Synchrotron
and inverse Compton emission from a population of high-energy
electrons and positrons (here collectively referred to as \lq\lq
electrons\rq\rq) distributed in a broken power-law spectrum. These
particles are presumably accelerated at, or close to, the termination shock (TS) of
the relativistic pulsar wind, which is located roughly where its momentum flux density
is balanced by the confining pressure of the external medium (for
young, isolated pulsars, the parent supernova remnant, for older ones, the
interstellar medium). The electron spectrum $N(\gamma)\propto\gamma^{-s}$ is
hard ($s\sim1.5$) at the low energies responsible for the radio to
optical emission, and softens ($s\gtrsim2.2$) at the high energies
responsible for the $X$-ray Synchrotron emission.  The X-ray morphology
suggests that electrons are preferentially accelerated in the
equatorial belt of an almost axisymmetric structure
\citep{kirklyubarskypetri09,amato14,porthetal14,olmietal15}.

In the best observed source --- the Crab Nebula --- the power-law index
at high energies is close to that predicted for first-order Fermi
acceleration at a parallel, ultra-relativistic shock front
\citep{bednarzostrowski98,kirketal00,achterbergetal01}. At first sight, this is 
surprising, since the magnetic field is not expected to be 
normal to the TS. On the contrary, the field is
embedded in the radial pulsar wind and is tightly wound up by the rotating neutron
star.  In an axisymmetric wind, this leads to a perpendicular magnetic
field configuration over almost the entire surface of the TS, and,
consequently, to strong suppression of the Fermi process
\citep{begelmankirk90, sironispitkovsky09, summerlinbaring12,sironikeshetlemoine15}.

Axisymmetry, however, is a good approximation only far from the
neutron star. Close to the surface of the star, the boundary
conditions impose a non-axisymmetric structure, which is frequently modeled as
an inclined magnetic dipole \citep[e.g.,][]{michel91}. In the region in which the wind is launched, this
structure is converted into a wave with an alternating magnetic field
in the equatorial zone
\citep{tchekhovskoyphilippovspitkovsky16}. Magnetic reconnection in
the equatorial zone was originally suggested by \citet{coroniti90} as a
way of energizing electrons in a \lq\lq striped wind\rq\rq.
Although subsequent work suggests that the process is too slow to annihilate
the field completely before the wind encounters the TS
\citep{lyubarskykirk01,kirkskjaeraasen03}, it nevertheless remains a promising
possibility, in particular for the production of gamma-ray pulses
\citep{mocholpetri15,ceruttiphilippovspitkovsky16} and flares
\citep{arons12,batypetrizenitani13,takamotopetribaty15}.

Therefore, given that reconnection is unlikely to be complete, the
equatorial zone of the wind arriving at the TS will contain a
substantial oscillating component of the magnetic field. Although
oriented perpendicular to the shock normal, this field is nevertheless
available for dissipation \citep{lyubarsky03}.  The dissipation
process itself can pictured in two scenarios, depending on the local value
of the plasma density:
\begin{enumerate}
\item 
\label{highdensitycase}
In a high-density plasma, the equations of MHD
can be expected to give a good description of the dynamics except
close to surfaces at which the polarity of the field reverses. In this case, the simplest picture
of the structure of the TS is one in which the magnitude of the magnetic field in the 
upstream plasma is constant, but its direction reverses at current sheets embedded in the flow.
At the TS, a fast-mode shock compresses the magnetized parts
as in the standard MHD picture. Because the plasma is strongly magnetized, the shock is weak, and 
does not significantly dissipate the magnetic energy \citep{kennelcoroniti84a}. 
However, the compression of the embedded current sheets triggers reconnection inside them, a phenomenon
referred to as \lq\lq driven reconnection\rq\rq.
As the sheets are advected downstream in the compressed, striped
pattern, dissipation becomes more and more important, causing them to 
expand and, ultimately, completely disrupt the pattern, thereby releasing the magnetic energy
\citep{petrilyubarsky07,sironispitkovsky11}.
\item
\label{lowdensitycase}
In a low-density plasma, the equations of MHD fail because of a lack
of charge-carrying particles.  This happens when the characteristic
scale on which the fields in the upstream plasma vary --- in this case
the rotation frequency of the pulsar --- is faster than the intrinsic
frequency of non-MHD waves, which is the local plasma frequency,
possibly modified by relativistic effects. The most important non-MHD
waves in this context are transverse, electromagnetic modes. In the
limit of low density, these correspond to vacuum waves that propagate
at $c$. At finite plasma density, their group speed is subluminal, but
their phase speed exceeds $c$, giving rise to the nomenclature
\lq\lq  superluminal waves\rq\rq
\citep{arkakirk12,mocholkirk13a,mocholkirk13b}. In this parameter
regime, the simplest picture of the structure of the TS is again one
in which the magnitude of the magnetic field in the upstream plasma is
constant, but, instead of reversing direction at a current sheet, the magnetic vector 
rotates smoothly in a monochromatic,
static shear. At the TS, this pattern converts into superluminal
waves, some of which propagate back into the upstream plasma, and
dissipate there in an \lq\lq electromagnetic precursor\rq\rq
\citep{amanokirk13}. 
\end{enumerate}
 
In a realistic system, the pulsar wind upstream of the TS will not
correspond exactly to either the striped case, or the monochromatic, static shear, so
that elements of each scenario may be present. However, their relative
importance will depend primarily on the ratio of the pulsar frequency to the
local plasma frequency. For almost all isolated pulsars, this is a large number at the TS
\citep{arkakirk12}. Therefore, we concentrate in this paper 
on particle acceleration in the low density scenario~\ref{lowdensitycase}.

First, we use the two-fluid code described in
\citet{amanokirk13} to generate a realization of the turbulent
electromagnetic fields in the proximity of the TS, and examine the
fate of test-particles moving in these fields. We find that a sizeable
fraction of these decouples from the fluid flow and is reflected,
thereby becoming available for further acceleration via the
first-order Fermi mechanism.  Assuming that small-scale magnetic
fluctuations scatter particles in both the upstream and downstream
plasma, we then use a Monte Carlo (MC) code to simulate this
acceleration process, in which particles cross and recross the
compound TS layer, consisting of an electromagnetic precursor and
shock compression. Despite the fact that the upstream plasma contains
an oscillating magnetic field oriented perpendicular to the shock
normal, we find that in the regime where superluminal waves propagate,
the spectral index of accelerated particles is close to that predicted
for a parallel shock, whereas, for high-energy particles, a softer
spectrum is predicted for the regime of driven reconnection.

The paper is organized as follows. In Sect.~\ref{2fluid} we describe
the two-fluid simulation, in \ref{test} the test-particle integration, 
and, in \ref{mc}, the Monte-Carlo simulations. The implications of
our results are discussed in Sect.~\ref{dis} and compared with 
results obtained in the high density, driven reconnection regime. 
Section~\ref{conc} summarizes our main conclusions.

\section{Two-fluid simulation}	\label{2fluid}

Numerical simulation of the structure of the TS in the low-density
regime presents a challenge.  Sophisticated, 3D-MHD simulations are
capable of addressing the global structure downstream of the TS and
the problem of dissipation in the pulsar wind nebula
\citep[e.g.,][]{porthetal14}, but these assume that all fluctuations
at the pulsar rotation frequency have been damped
away. Particle-in-cell simulations, on the other hand, are not limited
by this assumption, but they are computationally intensive. To date,
they have been performed in 2D and 3D, but only in the high-density regime,
where an extended, electromagnetic precursor does not form
\citep{sironispitkovsky11,takamotopetribaty15}. At present, the most
promising method available is two-fluid simulation, in which the electrons and positrons 
constitute separate, charged fluids that are coupled by Maxwell's equations 
\citep{zenitanietal09,barkovkomissarov16}.

\citet{amanokirk13} used a relativistic, two-fluid (electron and
positron) code (1D in space, 3D in velocity and electromagnetic
fields) to show that superluminal waves strongly modify the TS when
the ambient density is low enough to permit the propagation of
these waves.
They discussed in detail the dissipation processes
in the electromagnetic precursor upstream of the main compression, which
resembles a hydrodynamics sub-shock.
Whether or not this complex
structure is also capable of dissipating energy into a population of
accelerated, non-thermal particles extending over a broad range
in energy depends crucially on its ability to
reflect some of the particles that make up the incoming fluids
\citep[e.g.,][]{sironispitkovsky09}. To attack this question, we first
use the two-fluid code to generate an example of such a precursor
that has reached a long-lived, quasi-steady state.
\begin{figure*}[t!]
\begin{center}
\includegraphics[angle=270,scale=.6]{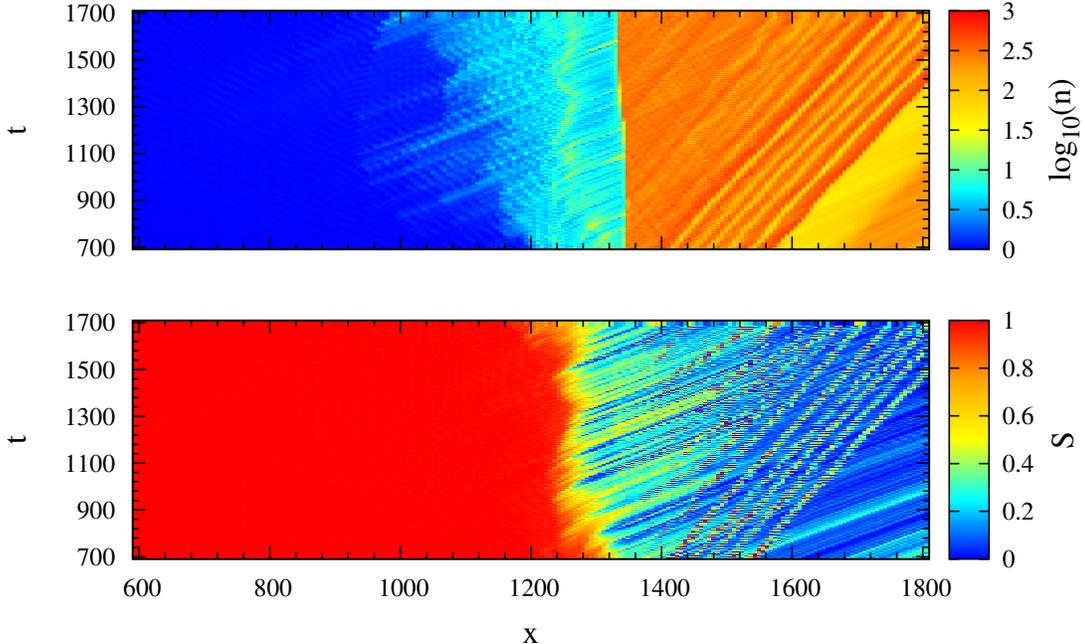}
\end{center}
\caption{Time evolution of the modified shock structure. The
  top and bottom panels show the proper density and the normalized
  Poynting flux profiles, respectively.  We show only the phase with a quasi-stationary
  precursor.\label{img:2fl}}
\end{figure*}

The set-up and parameters we choose are similar to those discussed by
\citet{amanokirk13}. A sinusoidal, circularly polarized, transverse magnetic shear
is assumed to be incident on the upstream boundary. 
In the following, we use the expressions \lq\lq shear wave\rq\rq
and \lq\lq striped wind\rq\rq interchangeably.
We define a frame of reference 
called the URF (upstream rest frame) in which the electric field of this wave
vanishes. Denoting quantities measured in this frame by a bar, 
and using cartesian coordinates with $x$ along the shock normal, its magnetic field 
is
\begin{equation}	\label{eq:by}
\overline{B}_{y}=+\overline{B}_{0}\cos(\overline{k}_{0}\overline{x})
\end{equation}
\begin{equation}	\label{eq:bz}
\overline{B}_{z}=-\overline{B}_{0}\sin(\overline{k}_{0}\overline{x})\,.
\end{equation}
The corresponding current is everywhere 
parallel to the magnetic field, whose magnitude is constant, 
so that the wave is, in 
fact, a force-free equilibrium. The proper density $n_0$ of 
the positron fluid equals
that of the electron fluid and is constant; the velocity 
$\overline{v}_0$ of the positron fluid, is parallel to $\overline{\bf B}$ 
and constant in magnitude (the electron fluid velocity is $-\overline{v}_0$).  

By a suitable choice of downstream boundary conditions, the reference
frame in which the simulation is performed 
(called the SRF --- shock/simulation rest frame)
is arranged to coincide
with that in which the shock-precursor structure is almost stationary. 
Seen from this frame, the URF moves along the $x$-axis at the \lq\lq shock speed\rq\rq\ $c\beta_{\rm
  s}$, which equals the phase speed of the incoming wave. Using
unadorned symbols for quantities measured in the SRF, the incoming
wave has wavenumber and frequency given by
$k=\Gamma_{\rm s}\overline{k}$, $\omega=c\beta_{\rm s}\Gamma_{\rm
  s}\overline{k}$, where the \lq\lq shock Lorentz factor\rq\rq\
$\Gamma_{\rm s}=1/\sqrt{1-\beta_{\rm s}^{2}}$.  Assuming the
pulsar moves slowly with respect to the TS, we identify the wave frequency with 
its rotation frequency.

Three dimensionless parameters are required to specify the initial
conditions for the simulation. In addition to $\Gamma_{\rm s}$, these
are the ratio $\Omega$ of the incoming wave frequency to the proper
plasma frequency associated with $n_0$: $\Omega=\omega/\omega_{{\rm
    p}0}$, $\omega_{{\rm p}0}=\left(8\pi n_0 e^2/m\right)^{1/2}$, and
the incoming magnetization parameter $\sigma_0$ (defined below).  Our
choice of $\Omega$ is restricted, on the one hand, to $\Omega>1$ by
the requirement that superluminal waves can propagate \citep[see
e.g.,][]{arkakirk12}. On the other hand, in order to find a precursor
that is contained within the simulation box, we find that
$\Omega\lesssim2$. Therefore, following \citet{amanokirk13}, we choose
$\Omega=1.2$ as a representative value.  In order to correspond to 
pulsar conditions, the shock Lorentz factor
$\Gamma_{\rm s}$ should be $\sim 10^4$ or larger,
but its value in the simulation is limited by technical issues associated with code stability,
and we adopt a compromise value of
$\Gamma_{\rm s}=100$. Finally, the magnetization parameter $\sigma$ is
defined in the simulation as the ratio of the fluxes in the $x$
direction of electromagnetic energy and particle enthalpy. However,
the requirement that the shock speed should greatly exceed the speed
of the fast magnetosonic wave: $\Gamma_{\rm s}\gg\sigma_0^{1/2}$,
implies that the velocity of the fluids in the URF is
non-relativistic:
$\overline{v}_0/c\approx\Omega\sigma_0^{1/2}/\Gamma_{\rm s}\ll1$, in
which case, $\sigma_0\approx\overline{B}_0^2/\left(8\pi
  n_0mc^2\right)$. In our simulation, we select $\sigma_{0}=25$, giving
$\overline{v}_0=0.06\,c$, and fix a cool initial fluid temperature
corresponding to a thermal velocity of $0.14\,c$.

\begin{figure}[t!]
\begin{center}
\includegraphics[angle=270,scale=.6]{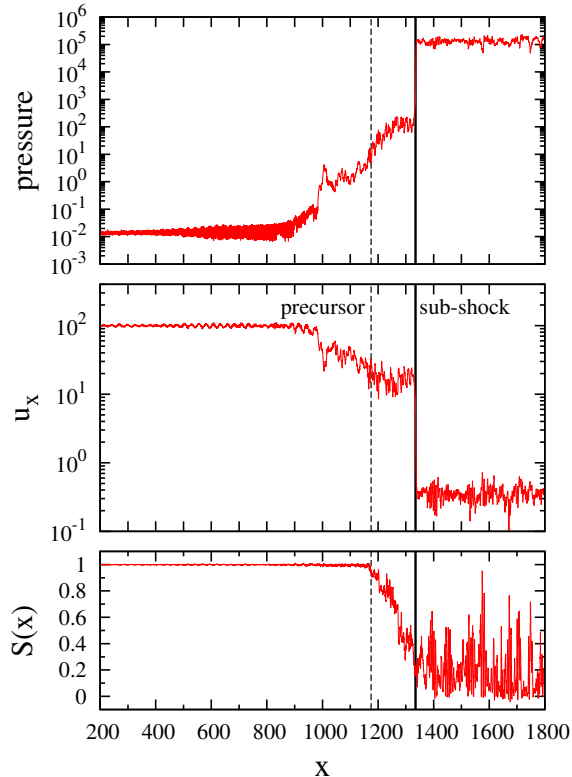}
\end{center}
\caption{Pressure (top panel), $x$ component of the plasma four-speed
  (middle panel) and normalized Poynting flux (bottom panel) profiles
  at $t=1700$. The solid and dashed black lines represent the position
  of the hydrodynamic sub-shock and leading edge of the precursor,
  respectively.\label{img:puts}}
\end{figure}

The resulting profiles of density and Poynting flux are shown in Fig.~\ref{img:2fl} 
where time and space are plotted in units of $1/\omega_{{\rm p}0}$ and 
$c/\omega_{{\rm p}0}$, respectively.
We confirm the findings of \citet{amanokirk13} 
that the breakout of the precursor is triggered by the launch
of superluminal waves which lead to the dissipation of the Poynting flux
$S$ carried by the incoming wave into enthalpy of the plasma. 
As can be seen in the top
and middle panels of Fig. ~\ref{img:puts}, where we plot the pressure
and the longitudinal component of the four-velocity of the plasma at
$t=1700$,
the dissipation is, at least in part, due
to the formation of small-scale shocks. These 
decelerate and heat the plasma
before it encounters the hydrodynamic sub-shock, which is located at $x_{\textrm{s}}=1335$ and
is represented by the black solid line in Fig.~\ref{img:puts}. 
In this figure, we depict the \lq\lq precursor\rq\rq\ as
that region located between the sub-shock and the point
$x_{\textrm{p}}=1175$, marked with a dashed black line,
where the bulk of the Poynting flux is dissipated.  
Here, there is a strong increase in plasma
temperature and pressure, which lowers the local plasma frequency, thereby permitting 
the propagation of low frequency superluminal modes. Further upstream,
the plasma is still perturbed by the propagation of superluminal waves, but 
dissipation is limited to increasing the plasma enthalpy at the expense of its
kinetic energy. 
This structure can be identified in Fig.~\ref{img:2fl}, by
comparing the extension of the region upstream of the shock where the
plasma density increases (the cyan region in the top panel) with the
region where the Poynting flux has its maximum value (red region in
the bottom panel). It can also be seen in fig.~\ref{img:puts}, 
by comparing the spatial profile of the Poynting flux with that of the 
plasma velocity at $t=1700$.  
In the precursor, the Poynting flux carried by the incoming wave is
almost completely converted into plasma enthalpy, as shown in the
bottom panels of both Fig.~\ref{img:2fl} and \ref{img:puts}, leaving a
region of turbulent fields downstream of the shock. Here,
the amplitude of the turbulence decreases as the distance to the shock increases, 
showing that the magnetic field of the incoming shear wave is completely annihilated.

\section{Test-particle trajectories}	\label{test}

To assess the ability of the TS to energize and potentially reflect
individual particles, we extract, from the simulation described above, 
the electromagnetic fields in a space-time volume in which a
quasi-stationary precursor is established, and follow the trajectories
of test particles in these fields.

Two technical issues arise in extracting the fields.  Firstly,
they are stored as vector quantities on a fixed grid in space-time.
In order to limit the data to a manageable volume, we keep the spatial
grid used in the two-fluid code (i.e., $4\times10^4$ points) and store
snapshots (typically 2000) taken at intervals of $\omega_{{\rm p}0}^{-1}$. 
Since we use an adaptive time-step to advance the
test-particle trajectories, it is necessary to interpolate the fields
on this two-dimensional grid. For this purpose, we use a 2-D cubic-spline 
algorithm \citep[see][]{Press86}, which ensures continuity and
differentiability of the interpolated function and its first
derivative, whilst avoiding spurious oscillations between grid points.

Secondly, examination of Fig.~\ref{img:2fl} shows that the precursor
can be regarded as quasi-stationary for times $t>700$. Thus, the total
time for which we are able to simulate this turbulence is limited to
$t\lesssim1000$. However, test particles can potentially spend a much
longer time in the precursor before being transmitted or
reflected. We deal with this problem by imposing periodic
boundaries in time on the fields, i.e., by folding as many times as necessary
the subset of snapshots between $t=700$ and $t=700+\tau$.
To check that this procedure does not introduce artifacts, 
we integrate several thousand trajectories, building the
electron spectrum and angular distribution
at both the upstream and downstream spatial boundaries for 
$\tau=250$, $500$, $1000$.

With this method, we obtain a representation of the spatial and
temporal dependence of the turbulent electric and magnetic fields in the
precursor, that covers fluctuations on timescales between
$\omega_{{\rm p}0}^{-1}$ and
$1000\omega_{{\rm p}0}^{-1}$, and length scales between $10^{-2}\,\lambda$ 
and the size of the computational box, roughly $500\lambda$.
These fields are responsible for scattering and pre-accelerating
super-thermal test-particles that are drawn from the electron and
positron fluids.

The test-particle integration procedure itself employs a standard fourth-order
Runge-Kutta algorithm, with the adaptive time-step routine described
by \citet{Press86}. This is applied to the equations of motion written
in the form given by \citet{kirkbellarka09}, but excluding the effects
of radiation reaction.  All trajectories are initiated far upstream of
the shock using one of two prescriptions: 
\begin{enumerate}
\item
\label{prescrone} The initial
four-momentum equals that of the electron fluid at that time-space
point. We use this prescription to study how and where the difference
between fluid and test particle trajectories manifests itself. 
\item
\label{prescrtwo}
The initial four momentum seen in the URF is randomly (isotropically)
distributed in angle and uniformly in Lorentz factor over the range
$[1:1.2]$. With this prescription test-electrons have roughly the same
energy as background electrons, but decouple more rapidly from the
fluid flow.  
\end{enumerate}
Each trajectory is followed until it terminates upon
reaching either the upstream or the downstream edge of the simulation
box.  The reflection probability is then the 
ratio of the number of trajectories that terminate at the
upstream absorbing boundary to the total number of trajectories.

\begin{figure}[!t]
\begin{center}
\includegraphics[width=0.3\textwidth,angle=270]{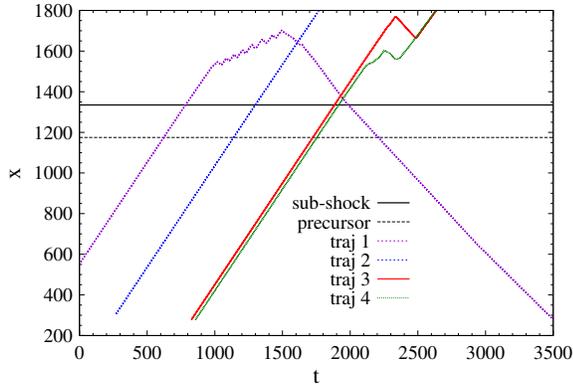}
\end{center}
\caption{Typical electron trajectories in the proximity of the sub-shock 
  (black solid line) and of its precursor (black dashed line). 
  Trajectories~1-3 are initiated as fluid elements (prescription~\ref{prescrone}). 
  Trajectory~4 is initiated as a test-particle (prescription \ref{prescrtwo}) at 
  the same location and time of trajectory~2 (for clarity plotted with 
  a small off-set in time).\label{img:traj}}
\end{figure}

Typical electron trajectories are plotted in Fig.~\ref{img:traj}, where the
black solid line represents the position of the sub-shock and the black
dashed line represents the approximate position of the leading edge of
the precursor. Trajectories~1-3 are initiated with the same
four-momentum as a local fluid element (prescription~\ref{prescrone}). Trajectory~4 
(for clarity plotted with a small off-set in time) starts with
$\overline{\gamma}=1.001$ and a random direction in the local
fluid frame (prescription~\ref{prescrtwo}) at the same
time and position as trajectory~3. 
Test electrons obey the same equations of motion as the fluid elements, except that 
the latter have an additional pressure force that prevents them intersecting each other. 
Consequently, test particles decouple from the background
plasma when the pressure term becomes appreciable. The fate of characteristic trajectories is 
shown in Fig.~\ref{img:traj}. Particles can travel across the shock with very little
deflection (trajectory~2) or undergo several reversals
(changes in the sign of $\beta_{x}$, trajectories 1, 3 and 4). If
they experience one or more reversals, electrons can subsequently be
registered at either one of the two spatial boundaries. We find that
the first reversal can only occur downstream of the sub-shock,
whereas the following can occur on either side of the sub-shock.
\begin{figure}[!h]
\begin{center}
\includegraphics[width=0.6\textwidth,angle=270]{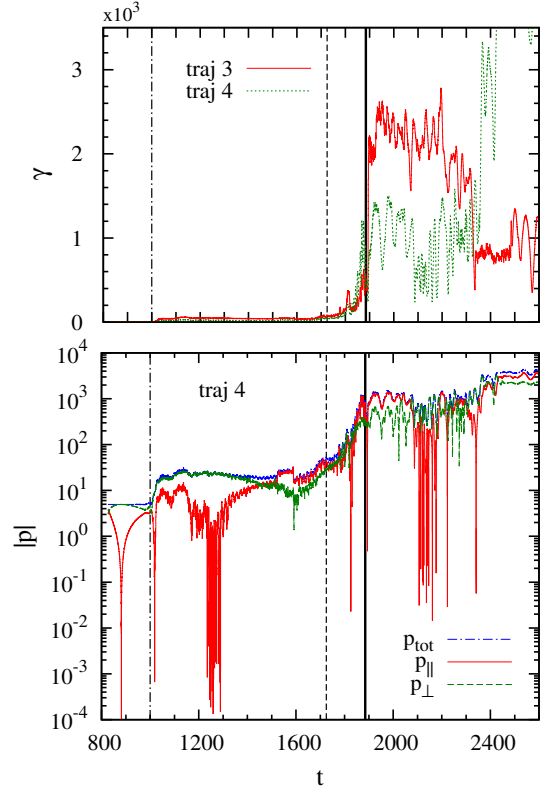}
\end{center}
\caption{Top panel: Comparison of the electron Lorentz factor expressed in the frame comoving with the fluid 
along $x$ for
trajectories 3 and 4. Bottom panel: Total (blue), longitudinal (red) and transverse (green) electron momentum for
trajectory 4 expressed in the same frame.
In both panels the vertical solid, dashed and dot-dashed black lines represent the sub-shock, precursor and superluminal
wave-front, respectively.\label{img:p}}
\end{figure}
In Fig.~\ref{img:p} we plot the electron Lorentz factor for
trajectories~3 and~4 (top panel) and the components of the electron
momentum (total (blue), longitudinal (red) and transverse (green)) for
trajectory~4 in a frame moving along the $x$-axis at the same speed as
the charged fluids (each of which has, to high accuracy, the same
$x$-velocity).  The vertical dot-dashed, dashed and solid black lines
represent the time at which an electron crosses the superluminal
wave-front, the leading edge of the precursor and the hydrodynamic
sub-shock, respectively. The top panel of Fig.~\ref{img:p} along with
Fig.~\ref{img:traj}, shows that there is no significant difference
between the trajectory and temporal behavior of the Lorentz factor
between the two prescriptions used to initiate the electron
trajectory. For both prescriptions, as seen in the frame comoving with
the fluids, test particles move with the plasma until they cross the
superluminal wave-front. There, the Lorentz factor slightly increases
and subsequently stays almost constant until the electrons reach the
leading edge of the precursor. In the precursor, the total energy of
test particles is enhanced by almost two orders of magnitude. When
electrons move downstream of the shock their Lorentz factor oscillates
about a constant value unless, as in the specific cases under
consideration (trajectories~3 and~4), more reversals occur, which can
cause a change in the total energy.  The bottom panel of
Fig.~\ref{img:p} shows that the transverse momentum of an electron
whose trajectory is initiated with prescription~\ref{prescrtwo} dominates over
its longitudinal momentum almost up to the leading edge of the
precursor\footnote{For an electron whose initial four-momentum equals
  that of the local fluid element (prescription~\ref{prescrone}), the initial transverse momentum is
  negligibly small and the longitudinal momentum dominates up to the
  wave-front.}. Close to the precursor the longitudinal momentum takes
over and dominates the energy balance for the rest of the
trajectory. Whether the energized electrons are transmitted to the
downstream absorbing boundary or reflected to the upstream one depends
on their parallel momentum and on the degree of turbulence they
encounter in the downstream. The reason for this is that if the
parallel momentum is large, or, conversely, the magnetic field they
encounter is small, their gyro-radius $r_{\textrm{g}}\propto
p_{\parallel}/B$ exceeds the size of the turbulent region and they
easily escape downstream.

\begin{figure}[t!]
\begin{center}
\includegraphics[angle=270,scale=.3]{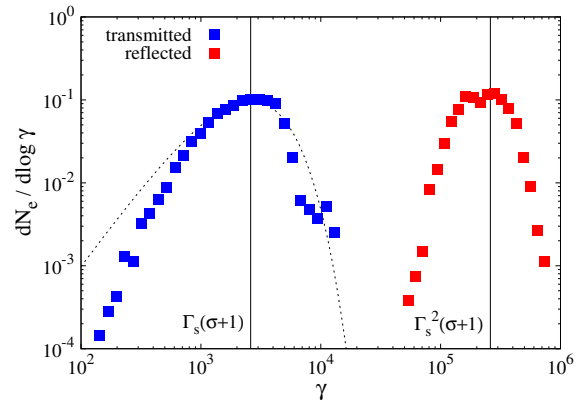}
\end{center}
\caption{Energy spectrum of test electrons transmitted (blue points)
  and reflected (red points) measured at the TS.  The spectrum of
  transmitted particles is expressed in DRF, whereas the spectrum of
  reflected particles is expressed in the URF. The black dotted line
  represents a relativistic Maxwellian peaked at
  $\Gamma_{\textrm{s}}(\sigma+1)$. Both spectra are normalized to the
  number of trajectories recorded at the related
  boundary.\label{img:spec}}
\end{figure}

As mentioned above, the ratio of the number of trajectories that
terminate at the upstream edge of the simulation box to the total
number simulated is interpreted as the injection/reflection
probability.  In order to check that this is accurate, i.e.,
represents a good estimate also for boundaries far removed from
the shock, we examine the particle fluxes across boundaries located at
different positions with respect to the shock front, both upstream
(referred to as UB) and downstream (referred to as DB).  For each pair
of boundaries, we compute the reflection and transmission
probabilities that would be found by treating these as absorbing
boundaries.
\begin{figure}[b!]
\begin{center}
\includegraphics[angle=270,scale=.3]{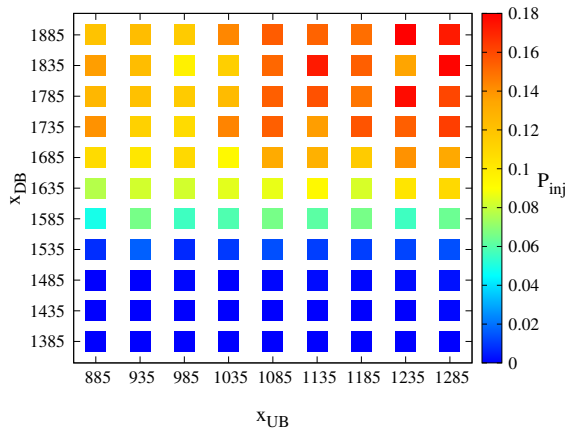}
\end{center}
\caption{Injection/reflection probability map as a function of the
  position of the upstream and downstream absorbing boundaries, UB and
  DB, respectively. The boundaries recede from the shock front moving
  leftwards on the UB-axis and upwards on the
  DB-axis.\label{img:refl}}
\end{figure}
The results are shown in Fig.~\ref{img:refl}, where the injection
probability is plotted as a function of the position of UB and DB. The
injection probability vanishes when DB is immediately downstream of
the shock (lowest row in the probability grid), since all the
trajectories are recorded at the boundary as soon as they cross the
shock front. When DB recedes from the shock (moving upwards in the
grid) the injection probability increases since more and more
electrons have the chance to be deflected into the upstream. On the
other hand, when UB is immediately upstream of the shock front, the
reflection probability is maximum (for a fixed position of DB), since
all the trajectories reflected upstream are registered as soon as they
reach UB. When UB recedes from the shock (moving leftwards in the
grid), the injection probability decreases since some of the
trajectories can be further deflected into the downstream by the
turbulent magnetic field in the precursor. The value of the injection
probability reaches an almost constant value when DB is sufficiently
far away from the shock and UB is upstream of the leading edge of the
precursor, indicating that an accurate estimate of the asymptotic
value has been reached.  For the simulation described in
Sect.~\ref{2fluid}, we find a reflection probability of
$P_{\textrm{inj}}\sim12\%$.  The spectra of transmitted (blue,
measured at the downstream edge of the simulation box) and reflected (red, measured at
the upstream edge of the simulation box) electrons is shown in Fig.~\ref{img:spec}. The
transmitted spectrum is plotted in the frame of reference computed
from the Rankine-Hugoniot conditions as corresponding to that in which
the downstream plasma would be at rest, if the electromagnetic fields
have been completely annihilated (the \lq\lq DRF\rq\rq), whereas the 
spectrum of reflected particles is expressed in the URF.
The spectrum of transmitted electrons resembles a relativistic
Maxwellian,
$M(\gamma)\propto\gamma\sqrt{\gamma^{2}-1}\,\textrm{exp}(-\gamma/\Delta\gamma)$,
shown, for comparison as a black dotted line.  This is in agreement
with the results of \citet{sironispitkovsky11} and
\citet{terakietal15}, who studied acceleration in the driven magnetic
reconnection and in the superluminal regimes, respectively. The peak
energy of the distribution is
$\Gamma_{\textrm{s}}(\sigma+1)$, shown in
Fig.~\ref{img:spec} by the leftmost vertical black solid line, which
means that in the electromagnetic precursor the electron energy
increases on average by a factor of $\sigma$, as expected for complete
dissipation of the Poynting flux. As for the spectrum of electrons
upstream, the average energy is
$\Gamma_{\textrm{s}}^{2}(\sigma+1)$, as expressed in the
URF. This energy value is represented by the rightmost vertical black
solid line in Fig.~\ref{img:spec}.  The fact that during the first
shock encounter the energy gain is $\propto\Gamma_{\textrm{s}}^{2}$ is well-known
\citep[e.g.,][]{vietri95,gallantachterberg99,achterbergetal01}. However,
here we show that the electromagnetically modified shock is able to
energize the particles by an extra factor of $\sigma$ due to the
dissipation of the magnetic field.  The spectrum of
reflected electrons in the URF is narrower than that of transmitted
electrons in the DRF.  In the URF, the angular distribution of
reflected electrons is strongly peaked about the direction
anti-parallel to the shock normal.  We
find that these results are insensitive to the period of the ensemble
of snapshots folded to integrate the test electron trajectories, and
have presented the results for the benchmark value
$\tau=1000$. Furthermore, the reflection probability map and particle
spectra are not sensitive to the injection prescription (1 or 2).

\section{Monte Carlo simulations} \label{mc}

In the idealized model of the pulsar wind as a magnetic shear,
electrons that escape across the boundaries of our simulation box that
encloses the TS shock are lost, in the sense that their trajectories
never return to the box. In a more realistic picture, however, the
incoming wave and the downstream plasma will contain magnetic
irregularities, such as Alfv\'{e}n waves or turbulence generated by
the escaping particles themselves, which can perturb these
trajectories.  The realization that this process leads to particle
acceleration underlies the theories of diffusive shock acceleration
at non-relativistic shocks as well as the corresponding (but
non-diffusive) process at relativistic shocks \citep[for introductory
reviews, see][]{drury83,kirkduffy99}.  This process is thought to be
suppressed at perpendicular, relativistic shocks 
\citep[e.g.,][]{summerlinbaring12,sironikeshetlemoine15}, but has so
far not been investigated for the case in which the upstream plasma
contains a field with reversing polarity, such as the magnetic shear
considered in the previous two sections.  For this purpose, we adapt a
well tried and tested Monte-Carlo (MC) technique, which assumes the
trajectories are stochastically perturbed by a scattering process that
causes diffusion of the direction of propagation, but does not change
the particle energy.
Our method is equivalent to that of \citet{summerlinbaring12}
in the limit of small-angle scattering; a recent implementation
can be found in \citet{takamotokirk15}.

At each time step, the particle momentum is advanced 
with an explicit first-order Euler's scheme \citep{achterbergkrulls92}. This
is done in the upstream plasma according to the
equations of motion in the unperturbed magnetic shear wave (Eqs.~\ref{eq:by} and \ref{eq:bz}),
whereas, in the downstream plasma, rectilinear motion is assumed, since the ambient fields vanish
according to the two-fluid simulations in Sect.~\ref{2fluid}.
Subsequently, a new direction of motion is randomly chosen in a cone of small
aperture $\delta\theta_{\textrm{max}}$ about the previous
direction, with a uniform distribution in the interval $[0:\delta\theta_{\textrm{max}}]$.
The aperture, combined with the time-step $\Delta t$, sets
the diffusion properties of the plasma which can be expressed as a
scattering length
\begin{equation}	\label{eq:lscat}
L_{\textrm{scat}}=\frac{6c}{\delta\theta_{\textrm{max}}^2/\Delta t}
\end{equation}
which is the length-scale over which a particle is, on average,
deflected by an angle $\pi/2$ by the magnetic
turbulence, and not by the ambient field.
The scattering length is a physical quantity, whose ratio to the
wavelength of the shear in the upstream plasma determines the physics of
acceleration. On the other hand, both $\Delta t$ and $\delta \theta_{\rm
max}$ 
are artificial quantities introduced by the discretization
procedure. For each simulation, we choose them to be sufficiently small
and independent of particle energy. This implies
that $L_{\rm scat}$ is also independent of particle energy, in contrast
with the simulations of, for example, \citet{summerlinbaring12}. We
discuss this aspect in more detail below.

Each trajectory is initialized at the \lq\lq shock front\rq\rq, which
corresponds to the upstream boundary of the two-fluid simulation, with
momentum directed along the shock normal into the upstream and with
$\gamma\gg\Gamma_{\textrm{s}}$.  Upon returning
to this boundary (which is fixed in the SRF), the Lorentz factor
$\overline{\gamma}$ is unchanged, but the corresponding quantity in
the SRF is larger. Using the test-particle integration technique
described in Sect.~\ref{test}, we find that particles re-entering the
TS with such high Lorentz factors have a negligible probability for
reflection --- they pass through the simulation box with no change in
Lorentz factor and remain essentially undeflected. Therefore, the MC
code picks up the trajectory as it emerges on the downstream side of
the TS with Lorentz factor and direction given by a Lorentz boost to
the DRF, and follows it until it either returns and recrosses the TS, or reaches a
boundary placed a fixed distance $d_{\textrm{abs}}$ downstream.  We
performed a series of tests on the position of this boundary, and
selected $500\,\widetilde{L}_{\textrm{scat}}$, the minimum value for
which the results showed no sensitivity. We denote by a tilde quantities
measured in the DRF. 

An important aspect of the simulations is the choice of
$\delta\theta_{\textrm{max}}$.  Electrons entering the upstream region
are restricted to a cone
$\overline{\mu}<-\beta_{\textrm{s}}$, where $\mu$ ($\overline{\mu}$) is the
cosine of the angle between the shock normal and the particle momentum
vector (and not the pitch-angle) in the SRF (URF).  This defines a small angle
\begin{equation}	\label{eq:thetac}
\sin\theta_\textrm{c}=\sqrt{1-{\beta}_{\textrm{s}}^{2}}=
1/\Gamma_{\textrm{s}}\sim\theta_{\textrm{c}}
\end{equation}
on which scale the angular distribution function of the particles is
expected to show structure, when viewed in the URF. 
As mentioned above, here we use this MC technique to describe
the process of diffusion in angle, which is represented in the transport
equation by a Fokker-Planck operator \citep[see e.g.,][]{kirkduffy99,
takamotokirk15}. Thus, in order to simulate diffusion in direction accurately
and to resolve the structures of the angular distribution, it is necessary to ensure
that $\delta\theta_{\textrm{max}}\ll\theta_{\textrm{c}}$. 

In contrast
to non-relativistic shocks, particles that cross into the downstream
region of the relativistic TS have a substantial probability of
escaping over the downstream boundary. To compensate for this, and
thereby minimize the effects of Poisson noise on the results, we
implement a particle splitting method. Every time an electron
completes five cycles (upstream-downstream-upstream), we use its
momentum as the initial conditions for $N$ daughter particles
($N=10$--$50$), each of which is given a statistical weight
$w_{\textrm{stat}}=1/N$ and is evolved independently.  

We record the Lorentz factor and angle $\cos^{-1}\mu$ at all shock
crossings for particles that have performed more than five
cycles. From this, we construct the asymptotic (high-energy) particle
distributions, averaged over the azimuthal angle:
$f(\gamma,\mu)\propto\gamma^{-s+2}g(\mu)$, and extract the dependence
on the polar angle $g(\mu)$ and the index $s$.

We have performed tests of our MC implementation for parallel shocks
(or, equivalently, those in which there is no ordered magnetic field),
and find good agreement between our results on the angular
distribution and spectrum of accelerated particles and those of
previous work \citep{kirketal00,achterbergetal01}, for a wide range of shock speeds,
${\Gamma}_{\textrm{s}}{\beta}_{\textrm{s}}=10^{-1}$--$10^{3}$.

In the case of the pulsar TS, we expect the level of turbulence
outside of the region simulated by the two-fluid code to be much
smaller than that inside it, and so restrict ourselves to upstream
scattering lengths that are longer than the wavelength of the magnetic
shear wave ($\overline{L}_{\textrm{scat}}=10^{3}\,\overline{\lambda}$
for regime I, and $\overline{L}_{\textrm{scat}}=10^{2}\,\overline{\lambda}$
for regime II, see below).
Downstream, there is no corresponding restriction, since the ordered
field is assumed to be completely dissipated in the precursor-TS structure.  

\begin{figure}[!t]
\includegraphics[width=0.6\textwidth,angle=270]{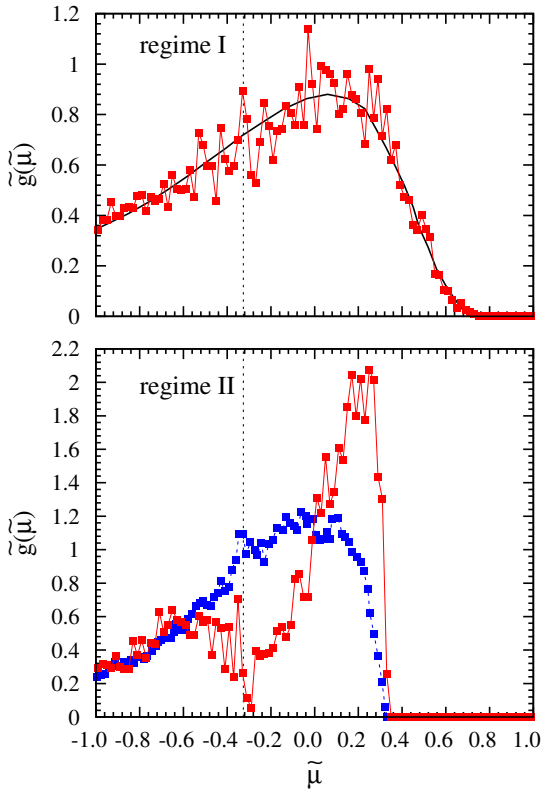}
\caption{Angular distribution expressed in the DRF for electrons
  accelerated at the pulsar wind TS ($\Gamma_{\textrm{s}}\beta_{\textrm{s}}=100$)
  in regime~I (red, upper panel) in comparison with pure scattering at relativistic shock
  \citep[black, eigenfunction method][]{kirketal00}, 
  and in regime~II (red, lower panel) in comparison with regular magnetic field deflection
  \citep[blue, numerical simulations,][]{achterbergetal01}. 
\label{img:ang}}
\end{figure}

In this case, our MC results show that electron acceleration proceeds
in two distinct regimes, dictated by the relative magnitude of the
wavelength of the stripes $\overline{\lambda}$ and of the electron
gyro-radius $\overline{r}_{\textrm{g}}$.  In the first --- regime I --- the gyro radius is large 
compared to the wavelength of the shear. This condition is automatically 
satisfied by the population of test particles depicted in Fig.~\ref{img:spec}.
In the second --- regime II --- the gyro radius of
accelerated particles is small compared to the wavelength of the
magnetic shear wave. This regime corresponds to the situation with driven
reconnection at the TS \citep{lyubarsky03,sironispitkovsky11}, 
rather than the scenario presented in Sects.~\ref{2fluid} and 
\ref{test}, and makes the implicit assumption
that also in this case,
particles can be injected into the relatively undisturbed upstream medium.

The angular distributions at the TS,
as expressed in the DRF, are plotted 
for $\Gamma_{\textrm{s}}\beta_{\textrm{s}}=100$ in
Fig.~\ref{img:ang}.  The top panel shows regime~I, and
compares our results (red) with those obtained for an unmagnetized (or
parallel) shock for $\Gamma_{\textrm{s}}\beta_{\textrm{s}}=10.0$
\citep[black, eigenfunction method,][]{kirketal00}. The bottom panel
shows regime~II, and compares our results (again in red) with
those obtained for a magnetized perpendicular shock  for
$\Gamma_{\textrm{s}}\beta_{\textrm{s}}=100.0$ \citep[blue,
numerical simulations,][]{achterbergetal01}.  The
vertical black dotted line represents the value of the speed of the
shock in DRF $\widetilde{\beta}_{\textrm{s}}$. This line divides
electrons crossing from downstream to upstream on the left-hand side
from electrons crossing from upstream to downstream on the right-hand
side. In regime~I, the wavelength of the magnetic shear is the
shortest relevant scale. As a result, electrons are unmagnetized and
scattering is the dominant source of particle deflection.  We find a
very good agreement between our results and the results of the
eigenfunction method for
$\Gamma_{\textrm{s}}\beta_{\textrm{s}}=10$ (which are indistinguishable from those of the
asymptotic solution at
$\Gamma_{\textrm{s}}\beta_{\textrm{s}}\to\infty$).  The angular
distribution is smooth, with a broad peak at
$\widetilde{\mu}\sim0.2$. The distribution extends to large values of
$\widetilde{\mu}$ for electrons crossing from upstream to
downstream. 
The
corresponding spectral slope is $s=2.23\pm0.01$ as expected for Fermi-like
acceleration at an unmagnetized relativistic shock front. 

\begin{table}[!b]
\begin{center}
\caption{Parameters of the simulation (including the energy range expressed in terms 
of particle gyro-radius) for electron acceleration at the pulsar wind TS 
($\Gamma_{\textrm{s}}\beta_{\textrm{s}}=100.0$) for the
scattering-dominated (I) and field-dominated (II) regimes.\label{tab:reg}}
\vspace{10pt}
\begin{tabular}{ccccc}
\tableline
\multicolumn{3}{c}{} \\
regime & $\overline{\Delta t}$ & $\overline{\delta\theta}_{\rm{max}}$ & energy range & $s$\\
\multicolumn{3}{c}{} \\
\tableline
\multicolumn{3}{c}{} \\
I & $10^{-6}$ & $3\times{10^{-6}}$ & $\overline{r}_{\textrm{g}}>30\overline{\lambda}$ & $2.23\pm0.01$\\
\multicolumn{3}{c}{} \\
II & $10^{-5}$ & $10^{-8}$ & $10^{-8}\overline{\lambda}<\overline{r}_{\textrm{g}}<10^{-1}\overline{\lambda}$
 & $2.65\pm0.03$\\
\multicolumn{3}{c}{} \\
\tableline
\end{tabular}
\end{center}
\end{table}

In regime~II, the gyro-radius of accelerating electrons is smaller
than the wavelength of the field. Consequently, the magnetic field in
the wind is the dominant source of deflection. Electrons crossing the
shock into the upstream medium essentially move into a uniform and
static magnetic field (as assumed in the simulations represented by
the blue curve) since these particles do not travel far enough into
the magnetic pattern to experience the rotation of the
magnetic field vector. However, the effect of 
the magnetic field of the shear wave
differs from that of a static and uniform field, as shown
in the bottom panel of Fig.~\ref{img:ang}, because in the former
the magnetic field at the shock front rotates while the electron
is engaged in an excursion downstream.
The number of electrons grazing the pulsar wind termination shock,
namely those with
$\widetilde{\mu}\sim\widetilde{\beta}_{\textrm{s}}$, is strongly
suppressed with respect to the uniform field case. Furthermore, the
distribution is strongly peaked at $\widetilde{\mu}\sim0.24$, while
the angular distribution obtained in the other case is smoother and
broader with a maximum at $\widetilde{\mu}\sim0.0$.  The resulting
spectral index is $s=2.65\pm0.03$, significantly different from the
case of regular magnetic field deflection
\citep[$s=2.28\pm0.01$,][]{achterbergetal01}.  The parameters of
these simulations are presented in Table~\ref{tab:reg}.

On physical grounds, one expects that the scattering length
$L_{\rm scat}$  
should increase as the particle's Lorentz factor grows,
whereas, according to our definition in Eq.~(\ref{eq:lscat}),
it is constant. For the acceleration timescale, and for the particle
energy spectrum in the transition region between regimes~I and II, this
is an important effect. However, since we are here concerned with the
angular distribution and spectrum only within these regimes, and do not
consider acceleration timescales, the chosen energy dependence of
$L_{\rm scat}$ does not play a role.

\section{Discussion}	\label{dis}

The two-fluid approach used by \citet{amanokirk13} allows one 
to investigate the
interaction between the pulsar wind and its termination shock
in a scenario which is complementary to that of driven magnetic reconnection
\citep[e.g.,][]{lyubarsky03,petrilyubarsky07,sironispitkovsky11}.
Superluminal waves mediate this interaction  
when the frequency of the wave exceeds the proper plasma frequency
($\Omega>1$ in our notation), whereas a combination of an MHD shock and
magnetic reconnection operates for $\Omega<1$, i.e., in a
relatively high density regime \citep[e.g.,][]{sironispitkovsky11}.

The breakout of the electromagnetic precursor due to the propagation
of superluminal waves triggers the dissipation of the Poynting flux of
the incoming wind, which
starts at the leading edge of the precursor, well 
upstream of the main compression, and proceeds almost to
completion, as illustrated in the bottom panels of Figs.~\ref{img:2fl}
and \ref{img:puts}. This mechanism creates a region of turbulent electromagnetic
fields where low frequency superluminal waves
can propagate and interact with the shear wave ahead of an essentially hydrodynamic sub-shock. 
The turbulent region is effective in
energizing the electrons carried along with the wind, causing 
a sizeable fraction of them to be reflected after crossing the sub-shock. 
Upstream of the leading edge of the precursor
the incoming plasma is still perturbed, most likely because of the propagation
of high frequency superluminal waves, but there is only limited 
dissipation of the Poynting flux.  
The test-particle approach used here shows that
electrons decoupled from the background plasma gain less than $1\%$ of their final energy
in this region. This is due to
the presence of a small electric field in the frame comoving with the
electron-positron plasma.  This heats and compresses the background plasma.
On the other hand, 
test-particles gain momentum in the
transverse plane, as shown in the bottom panel of
Fig.~\ref{img:p}. Since the electric field is perpendicular to the
magnetic field, electrons acquire a small component of momentum in the
transverse plane perpendicular to the magnetic field and start
gyrating about the local magnetic field line. 
This causes the \lq\lq bouncing\rq\rq pattern of the parallel momentum
(red) in the bottom panel of Fig.~\ref{img:p}, where the peaks are due to
motion parallel and anti-parallel to the $x$-axis.  
In the precursor, the net electric field in the comoving frame
grows substantially, and electrons are accelerated in the direction
perpendicular to the bulk motion by a non-MHD electric field arising
in the interaction between the superluminal wave and the incoming
shear wave \citep[see][for a discussion on non-MHD fields in the
two-fluid simulation]{amanokirk13}. The
parallel momentum can either increase (as for trajectory 4) or
decrease according to the phase of the gyration when the electron
enters the precursor. We stress that in
Figs.~\ref{img:puts}-\ref{img:p}, the black dashed line represents the
average position of the leading edge of the precursor which is
subjected to small fluctuations over the simulated time frame.  Thus,
the energy increase associated with the electron entering the precursor
region can occur slightly upstream or downstream of the line depicted in
the plots.

Even though the precursor is very turbulent, only the 
electromagnetic fields that survive downstream of the sub-shock are able to cause the
first reversal of the electron trajectory (change of the sign of
$\beta_{x}$) and turn trajectories around towards the upstream. This is 
illustrated by the sample of trajectories plotted
in Fig.~\ref{img:traj} and holds true for all of the several
thousand test-particle trajectories we have examined. 

We find an asymptotic value of the reflection probability at the TS in
a striped wind (obtained
for boundaries far from the sub-shock) of $P_{\textrm{inj}}\sim12\%$,
which is close to that found for parallel shocks using MC techniques
\citep{bednarzostrowski99,achterbergetal01}.
This implies that electrons are injected into a subsequent, first-order Fermi
process with comparable efficiency in the two cases, despite the fact that
a striped wind requires particles 
to be boosted in energy before they can be reflected.

The spectrum of test electrons downstream of the shock resembles
a relativistic Maxwellian peaked at $\sim\sigma\Gamma$, close to that 
expected for complete dissipation of the wave magnetization
into test particles. This resembles the high 
density regime, where magnetic reconnection 
is expected to accelerate particles to a similar energy 
\citep{kirk04,kaganetal15,uzdensky16}. 
However, the dissipation of the magnetic field
at the wind TS is not sufficient in itself to accelerate
electrons into the power-law spectrum required to explain the
observations of PWNe, which requires an additional acceleration mechanism. 

The first-order Fermi process provides a possible scenario for this mechanism
in the equatorial region of the TS in PWNe.  The acceleration regime
is determined by the size of the gyro-radius
of the electron in
comparison with the wavelength of the shear wave, or, in other words,
of the wavelength of the stripes in the pulsar wind.

If $\overline{r}_{\textrm{g}}\gg\overline{\lambda}$, particles only perform
a partial gyration about the magnetic field line (as seen in URF) before
encountering the opposite phase of the
magnetic shear wave and gyrating in the opposite direction. In this
case, the variation of $\overline{\mu}$ during the partial orbit
(electrons are confined into a cone $\overline{\mu}<-\beta_{\textrm{s}}$
upon entering the upstream region, see Sect.~\ref{mc}) is not
sufficient for the shock front to overcome the particle. This result
was already suggested by our test-particle simulation, where electrons
travelling upstream of the leading edge of the precursor in the
direction anti-parallel to the wind are always recorded at the
upstream absorbing boundary.  In regime~I, the magnetic field is
unimportant and only scattering off magnetic turbulence can provide
the deflection necessary for the shock to overcome the electron
(scattering-dominated regime).
Regime I applies as long as the magnetic field is unable to
provide the necessary deflection for electrons to cross the shock
for the subsequent excursion in the downstream region. Consequently, 
the magnetic field is irrelevant for the acceleration process for 
\begin{equation}	\label{eq:sb}
\frac{\overline{\lambda}}{\overline{r}_{\textrm{g}}}<\theta_{\textrm{c}}\,.
\end{equation}
For electrons reflected at TS in our test-particle approach,
this condition translates to 
\begin{equation}
\overline{\gamma}>
2\pi\sqrt{\sigma_0}\Gamma_{\textrm{s}}^{2}/\Omega
\end{equation}
which is automatically satisfied for the majority of reflected
particles, since these have
$\overline{\gamma}\sim\sigma_0\Gamma_{\textrm{s}}^{2}$, and in this regime
$\Omega\sqrt{\sigma_0}>1$.
Thus, under pulsar wind conditions, although the
upstream plasma is highly magnetized, the equatorial section of TS
acts as an unmagnetized relativistic shock, producing the angular
distribution and the power-law spectral index characteristic of
first-order Fermi acceleration at unmagnetized relativistic shock fronts.

The second acceleration regime found in our MC approach (regime~II,
field-dominated) applies when Eq.~(\ref{eq:sb}) is not satisfied. This
would require injection of particles in the upstream with
$\overline{r}_{\textrm{g}}\ll\overline{\lambda}$, at odds with the
results of Sect.~\ref{test}. Assuming that such an injection mechanism
can be provided by another mechanism, such as reconnection, the
trajectory which enters the upstream is, nevertheless, bound to a
specific field line which drives it back towards the shock. As seen in
the URF, the electron again performs only a partial gyration about the
field line.  Since the shock is highly relativistic, it overruns the
electrons soon after the condition
$\overline{\mu}<-\beta_{\textrm{s}}$ is met, and prevents them from
acquiring a large deflection.  Therefore, the resulting angular
distribution, which is plotted in the bottom panel of
Fig.~\ref{img:ang}, shows almost no particles with
$\widetilde{\mu}>0.4$.  In addition, the number of electrons grazing
the shock and accumulating at $\widetilde{\mu}=-0.33$
is suppressed in this case in comparison with the uniform magnetic
field case (blue curve in the bottom panel of Fig.~\ref{img:ang}).  
In a uniform field, electrons moving
parallel to the shock front ($\mu\sim0$) and to the magnetic field in
the upstream which perform very short excursions downstream (namely
with little variation of $\mu$) are basically \lq\lq bound\rq\rq to
the magnetic field line upstream. These electrons cross the shock
multiple times (with very little energy gain at each cycle) and
generate many shock crossing events with
$\widetilde{\mu}\sim-\widetilde{\beta}_{\textrm{s}}$ in our plot.
Such trajectories are absent in the case of a shear wave.  In fact,
during an excursion downstream, the orientation of the upstream field
at the shock front changes and trajectories that return to the shock
can suffer a very large change in pitch angle.  This increases the
average upstream excursion time and shifts the peak of the
distribution of particles crossing the shock from upstream to
downstream to larger values of $\widetilde{\mu}$.  The larger average
deflection upon returning to the shock leads to a
larger average energy gain, but also to an increased escape
probability.  As a result, the spectrum of particles accelerated at a
shear wave in regime~II is softer than that of particles accelerated
in a uniform field.

Comparing the two regimes we note that magnetic dissipation produces a
Maxwellian spectrum in regime~I, (as noted in Sect.~\ref{test}),
whereas regime~II (driven reconnection) produces a power-law spectrum
with a relatively hard $s\sim1.5$ spectrum for $\gamma\ll\sigma_{0}\Gamma_{\textrm{s}}$
up to a cut-off
\citep{sironispitkovsky11}.  In principle, the different signatures of
acceleration contained in the angular distributions and, more
importantly, in the energy spectra can be used to discriminate between
the dissipation mechanisms that operate in the proximity of the TS.
We stress that our conclusions apply only in the equatorial region of
the pulsar wind TS, where the magnetic field averaged over the
wavelength of the stripes is small.  At higher latitudes, Fermi-type
acceleration is inhibited by the large non-oscillating component of
the magnetic field, which advects electrons away from the shock,
quenching the acceleration process.

\section{Conclusions}	\label{conc}

The relativistic shock front terminating the striped wind emitted by a
pulsar in the equatorial region acts as an effectively unmagnetized
shock. Electrons and positrons accelerated at this location have an
angular distribution and a power-law spectral index consistent with the
predictions of first-order Fermi acceleration at parallel, relativistic shocks.

We have used a combination of two-fluid, test-particle and Monte Carlo
simulations to investigate electron acceleration at the termination
shock in PWNe. The two-fluid simulations provide a representation of
the turbulent electromagnetic fields in the proximity of the shock,
capturing the process of dissipation of the magnetic field in the
stripes of the wind and the formation of an electromagnetic
precursor. We have shown that test-particle electrons propagating in
the precursor increase their energy on average by a factor of
$\sim\sigma_0$, as expected for almost complete dissipation of the
Poynting flux at the TS. These particles form a relativistic
Maxwell-like distribution peaked at $\sigma\Gamma_{\textrm{s}}$ in the
downstream of the shock. A fraction $P_{\textrm{inj}}\sim12\%$ of the
incoming electrons is reflected upstream, forming a population of
particles available for further acceleration.  We have shown that
subsequent stochastic acceleration at the shock can operate in two
regimes. These are scattering- or field-dominated, according to the
relative magnitude of the electron gyro-radius and wavelength of the
stripes. Since, in pulsar wind environments the wavelength of the
stripes is always the smallest relevant length-scale, the acceleration
is likely to proceed in the scattering-dominated regime
(regime~I). The resulting angular distribution and spectral index
$s=2.23\pm0.01$ are similar to those obtained for Fermi acceleration
at relativistic and unmagnetized shocks (when radiation losses are
neglected).  Interestingly, this slope is very close to the that
needed to explain the TeV emission from the Crab.
The field-dominated regime (regime~II) corresponds to acceleration in
driven magnetic reconnection and produces a softer spectrum of slope
$s=2.65\pm0.03$. However, this regime requires a different injection mechanism,
which we do not discuss in this paper.

Finally, although we discuss our results in the context of PWNe, they
may also be relevant to other sources that contain relativistic,
magnetically dominated flows, such as AGNs or GRBs.

\acknowledgments

We thank T. Amano for placing the two-fluid code at our
disposal and for indispensable help and advice, and M. Takamoto for
many inspiring discussions. We also thank the anonymous referee
for insightful comments and suggestions which improved the quality
of our manuscript. S.G. acknowledges support from the
International Max-Planck School for Astronomy and Cosmic Physics.

\bibliographystyle{apj}
\bibliography{paper_bib}

\clearpage

\end{document}